\documentclass[prb,twocolumn,superscriptaddress,showpacs]{revtex4-1}
\usepackage{graphicx}
\usepackage{amsmath}
\usepackage{amsfonts}
\usepackage{float}
\usepackage[colorlinks,linkcolor=blue,citecolor=blue,urlcolor=blue]{hyperref} 
\usepackage{natbib}
\newcommand{\RomanNumeralCaps}[1]
    {\MakeUppercase{\romannumeral #1}}

\begin{document}
\title{Analytical solutions of inhomogeneous transverse field Ising models}

\author{ Abhijit P. Chaudhari}
 \email{apravin.chaudhari.phy16@iitbhu.ac.in}
 \affiliation{Department of Physics, Indian Institute of Technology (Banaras Hindu University), Varanasi - 221005, India}

\author{Rajeev Singh}
\affiliation{Department of Physics, Indian Institute of Technology (Banaras Hindu University), Varanasi - 221005, India}

\author{ Sunil~K.~Mishra}
 \email{sunilkm.app@iitbhu.ac.in}
 \affiliation{Department of Physics, Indian Institute of Technology (Banaras Hindu University), Varanasi - 221005, India}
\date{\today}

\begin{abstract}
The inhomogeneous transverse field Ising models mainly impurity based and the joint chain
are analysed analytically using Jordan-Wigner transformations. The effects of 
inhomogeneities on the phase transition have been discussed in detail. We constructed 
an ansatz to diagonalize
the two models which are taken into consideration. The inhomogeneity 
is quantified by a coupling parameter, which can be tuned to control the 
occurrence of quantum phase transition in these models. We have shown
a systematic setup using which we can 
generalise the solution to a system with an arbitrary number of impurity
sites and junctions, which are separated by at least two lattice sites. We have 
analysed the quantum critical point by calculating the correlation functions, 
transverse magnetization and the gap between the ground state and the 
first excited state.  
\end{abstract}

\pacs{03.67.Hk, 75.85.+t, 75.10.Pq}
\maketitle
\section{Introduction}
Quantum phase transition (QPT) is one of the most intricate and interesting phenomena which emerges in quantum many-body physics. \cite{Vojta2003,Eriksson2008,Zhong2010,Islam2011,Sachdev2000,RevModPhys.69.315,sachdev_2011} Their manifestation in quantum many-body systems is realised when the ground state of the system undergoes a qualitative change by tuning a non-thermal parameter (say transverse magnetic field ($h$) for instance) across its critical value ($h_c$). Unlike the thermally induced phase transitions, a quantum phase transition is triggered by quantum fluctuations which can take place even at the absolute zero temperature. The ground state properties of a system is determined by its intricate structure. The degree of complexity of this structure is captured by the entanglement that pervades ground state of the system.\cite{Latorre2003,PhysRev.108.1175,PhysRevLett.50.1395,Stelmachovic2004,Peschel2009,Calabrese2016,Naik2018} As a consequence of quantum phase transition, the entanglement structure of the ground state changes too, which gives us an opportunity to engineer quantum systems where the entanglement can be manipulated as per our will.\cite{Bayat2017,Gu2003,Campbell2011,Its2005a,Vidal2003,Clark2010,Osterloh,Sengupta2009,Osborne2002,Amico2008} Such a viewpoint is important in recent times where entanglement is seen as a crucial resource to process and send information using various protocols like superdense coding, quantum teleportation etc.\cite{nielsen_chuang_2010,PhysRevA.54.3824,PhysRevA.53.2046,Bennett2000,Venuti2007}

Quantum spin chains are useful systems for theoretical and experimental investigations because of the accessibility of QPT. Various types of quantum spin chain systems have demonstrated to be effectiveness for teleportation and direct state transfer protocols.\cite{Bose2007,Bose2003,DiFranco2008,Lorenzo2015,Lorenzo2013,0295-5075-119-3-30001} Apart from the state transfer applications, quantum spin chains have also found applications in hardware development of quantum computer.\cite{Marchukov2016,PhysRevA.87.062309,PhysRevA.77.050306}

 Interacting quantum spin system are interesting topics of study also because of the extensive body of work performed in the area of exactly solvable models. There are various analytical and numerical techniques for the investigation of statistical properties of these models.\cite{Quasi1994,Adegoke2010,RevModPhys.80.517,Eisert2010,Peschel2012,DeChiara2008} Recently after the quantum Newton's cradle experiment was reported, the scientific community has started taking interest in physically realising these models because of their exotic behaviour against the quantum non-integrable models.\cite{Kinoshita2006}The models that we consider in this work also come under the same category.
 
 In realistic situations the impurities are always present in the system and they recast drastically to the various properties of the system. For instance in the case of a metal, the magnetic impurity leads to a well studied phenomena known as Kondo effect.\cite{Schrieffer} The impurity in a spin chain adds the inhomogenity in the system and modifies the exchange interaction strength locally.\cite{Apollaro2017} The broken translational symmetry due to the imurity sites in these system implies that the state of the system becomes no more invariant under this symmetry and lead to  major change in the ground state properties of the realistic system as compared to the pure system.   
  In the case of impurity based transverse field Ising model (TFIM), we consider an impurity located at the centre of TFIM. This model that we consider can be realised as the physical description of isolated impurities in solid state systems. It can also be used for modeling the random noise between two quantum information channels. In another model we consider the junction of two TFIMs with different coupling strength. This model can be realised as a physical description of junction between two quasi one dimensional solid state systems. It can be used for modeling the junction of two quantum information channels with different coupling strengths.
 
 Development of ultra-cold atom simulations have allowed us to practically realise these models on optical lattices.\cite{Simon2011} It is straightforward to show that the creation and annihilation operators describing the hardcore bosons are algebraically identical to lowering and raising operators of spin systems. Hence the two models that we considered can also be simulated using ultra-cold atoms. The inhomogeneity coupling strength turns out to be the tunnelling coefficient for the ultra cold atoms on the optical lattice which can be controlled very precisely in a properly engineered simulation. The paper is organised as follows: In Sec.\ref{Sec. 2} we introduce the general inhomogeneous transverse field Ising models and also discuss the conventional method used for diagonalising these models. In Sec.\ref{Sec. 3} and Sec.\ref{Sec. 4} we discuss the analytical solutions for impurity based TFIM and joint-chain TFIM respectively and the salient features of the solutions. In Sec.\ref{Sec. 6}  we discuss a systematic way of developing the ansatz that was used for analytically diagonalising the two models.  In Sec.\ref{Sec. 7} we present a study of quantum phase transition which is observed in these models. Finally in Sec.\ref{Sec. 8} we draw our conclusions.
 
 \section{Inhomogeneous transverse field Ising models}
 \label{Sec. 2}
We consider a class of inhomogeneous transverse field Ising models\cite{PhysRevB.60.4195,0305-4470-23-13-037,PhysRevB.57.11404,1751-8121-45-9-095002,PhysRevLett.113.076401,Tsvelik2013,1367-2630-16-3-033003,Affleck2008,Apollaro2017} defined by the Hamiltonian
\begin{equation}
 \hat{H} = -\sum_{i=1}^{N-1}J_iS_{i}^{x}S_{i+1}^{x} - \sum_{i=1}^{N}S_{i}^{z}. 
 \label{ham1}
 \end{equation}
The inhomogeneity results from different values of $J_i$'s. $S_i^{\alpha} = \sigma_{i}^{\alpha}$, where $ \sigma_i^{\alpha} (\alpha = x , y , z)$ are the Pauli matrices corresponding to the components of spin$-\frac{1}{2}$ quantum operator at $i^{th}$ site. It is a well known fact that gobal spin flip symmetry ($U_{PF}$) is a symmetry of the Hamiltonian mentioned above, where $U_{PF} = \prod_{i=1}^{N}\sigma_i^{z}$. In this study we consider two different models. Model \RomanNumeralCaps 1 (impurity based TFIM) and  Model \RomanNumeralCaps 2 (joint-chain transverse field Ising model). We will present the analytical solutions for  these models and use them to find the signatures of quantum phase transition in these systems. 
 \\ \\
 We use Jordan-Wigner transformation\cite{Nielsen2005} to map the spin$-\frac{1}{2}$ operators onto the lattice
fermionic operators (${a_i, a_i^{\dagger}}$), which gives us $S_i^xS_{i+1}^x = (a_i^{\dagger}-a_i)(a_{i+1}+a_{i+1}^{\dagger})$ and  $S_i^{z} = [ a_i , a_i^{\dagger}]$. As a result we obtain a quadratic Hamiltonian in fermionic operators. By introducing Bogoliubov-de Gennes quasiparticle operators ({${b_{j}, b_{j}^{\dagger}}$} ) (where  $1\leq j \leq N$) which are related to (${a_i, a_i^{\dagger}}$) by Bogoliubov transformations 
\begin{align}
b_{j} &= \sum_{i=1}^{N}\frac{\phi_{ji} + \psi_{ji}}{2}a_i + \frac{\phi_{ji}-\psi_{ji}}{2}a_i^{\dagger} \nonumber, \\
b_{j}^{\dagger} &= \sum_{i=1}^{N}\frac{\phi_{ji} + \psi_{ji}}{2}a_i^{\dagger} + \frac{\phi_{ji}-\psi_{ji}}{2}a_i,
\end{align}
we can recast the Hamiltonian $\hat{H}$ in the form 
\begin{equation}
\hat{H} =- \sum_{j=1}^{N}2\Lambda_{j}b_{j}^{\dagger}b_{j} + \sum_{j=1}^{N}\Lambda_{j}.
\end{equation}
Here $\phi_{j}$ and $\psi_{j}$ are considered to as $N$ component vectors, which are real solutions for following matrix equations and $\Lambda_j^2$ are the real and positive eigenvalues of the matrix $(\alpha+\beta)(\alpha-\beta)$ \cite{Lieb1961}.
 \begin{align} 
(\alpha+\beta)(\alpha-\beta)\phi_{j}^{T} &= \Lambda_{j}^2\phi_{j}^{T} \label{eq:Matrixequation1},\\
(\alpha-\beta) \phi_{j}^{T} &= \Lambda_{j}\psi_{j}^T \label{eq:Matrixequation2},\\
(\alpha+\beta) \psi_{j}^{T} &= \Lambda_{j}\phi_{j}^T \label{eq:Matrixequation3}.
\end{align} 
\begin{figure}[!t]
\begin{center}
\includegraphics[width = \linewidth]{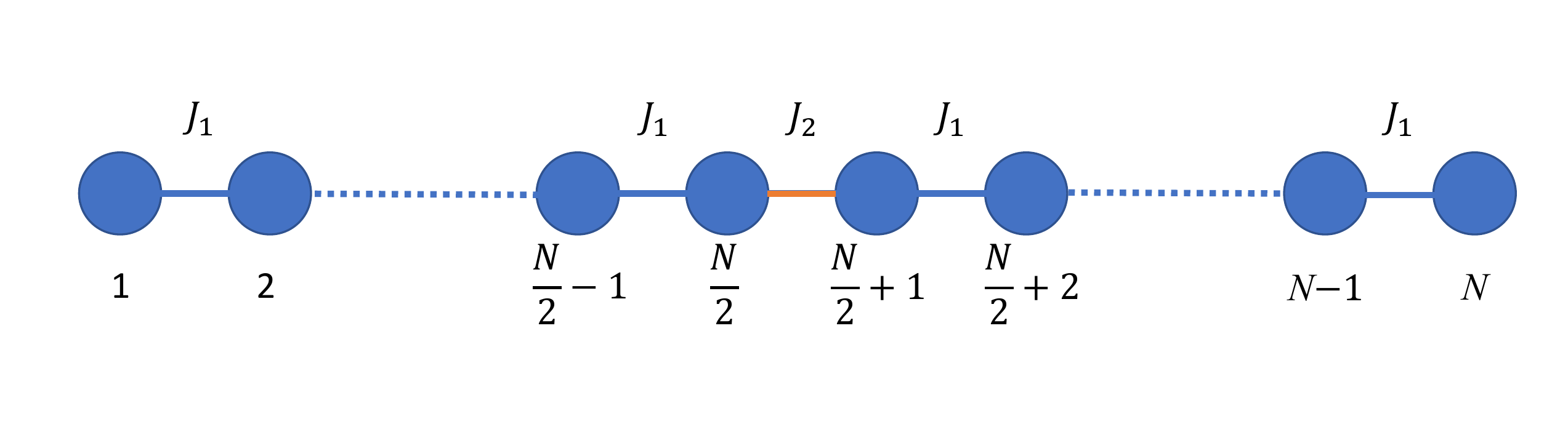}
\caption{Schematic for the impurity based TFIM. The presence of impurity modified the 
exchange coefficient between $\frac{N}{2}$ and $\frac{N}{2}+1$ sites on the spin chain.  The red coloured link with exchange interaction strength $J_2$ differentiated from the other blue coloured bonds. Even system size and open boundary condition are considered.}
\label{impurity_fig}
\end{center}
\end{figure}

Here $\phi^T$ represents the transpose of the matrix $\phi$. The only objects that change when we move from one model to another are the $\alpha$ and $\beta$ matrices ($\alpha_{jk}$ is the coefficient of $a^{\dagger}_ja_k$ and $\beta_{jk}$ is the coefficient of $a_ja_k$ in the Hamiltonian obtained after Jordan-Wigner transformation). These matrices contain the information which characterizes a given model. As a result, the problem reduces to solving the matrix equations \eqref{eq:Matrixequation1},\eqref{eq:Matrixequation2} and \eqref{eq:Matrixequation3}. 

\section{Model I : Impurity Based Transverse Field Ising Model} \label{Sec. 3}
Let us first consider a one dimensional transverse field Ising model with an impurity present at the center
of the spin-chain with open boundaries.\cite{Uzelac1981}
We call this model as model \RomanNumeralCaps 1 (impurity based TFIM) and represent the impurity by modifying the coupling parameter to $J_2$ between the $\frac{N}{2}$ and $\frac{N}{2}+1$ sites on the spin chain  ($N$ is considered to be even ). $J_1$ is the coupling parameter along the host chain on which the impurity is embedded (see fig. \ref{impurity_fig}). $h$ represents the transverse magnetic field at each site. The nearest neighbour
coupling parameter between the $i^{\rm{th}}$ and $(i+1)^{\rm{th}}$ site is given as 
 \begin{equation}
 J_i = (J_1/h)(1-\delta_{i,\frac{N}{2}}) + (J_2/h)\delta_{i,\frac{N}{2}}.
\end{equation}
 For this choice of $J_i$ in eq. (\ref{ham1}) we can proceed with the steps discussed in the previous section and write the component form of the matrix eq. \eqref{eq:Matrixequation1} for model \RomanNumeralCaps 1, is
\begin{equation}
\phi_{(i-1)j}^{T} + \phi_{(i+1)j}^{T} = \biggr(\frac{J_1^2+h^2-h^2\Lambda_j^2}{hJ_1}\biggr)\phi_{ij}^{T},  \end{equation} 
where $1 \leq i \leq (\frac{N}{2}-1)$ and $(\frac{N}{2}+2) \leq i \leq (N-1)$. The initial condition for equation \eqref{eq:Matrixequation1} is 
\begin{align}
\phi_{2j}^{T} &=\biggr(\frac{h^2-h^2\Lambda_j^2}{hJ_1}\biggr)\phi_{1j}^{T},
\end{align}
and the boundary conditions are:
\begin{align}
\phi_{(\frac{N}{2}-1)j}^{T} +\frac{J_2}{J_1}\phi_{(\frac{N}{2}+1)j}^{T} &= \biggr(\frac{J_1^2+h^2-h^2\Lambda_j^2}{hJ_1}\biggr)\phi_{\frac{N}{2}j}^{T},\\
\frac{J_2}{J_1}\phi_{\frac{N}{2}j}^{T}+\phi_{(\frac{N}{2}+2)j}^{T} &= \biggr(\frac{J_2^2+h^2-h^2\Lambda_j^2}{hJ_1}\biggr)\phi_{(\frac{N}{2}+1)j}^{T},\\
\phi_{(N-1)j}^{T} &= \biggr(\frac{J_1^2+h^2-h^2\Lambda_j^2}{hJ_1}\biggr)\phi_{Nj}^{T}.
\end{align}
\begin{figure}[!t]
\begin{center}
\includegraphics[width = 1.0\linewidth]{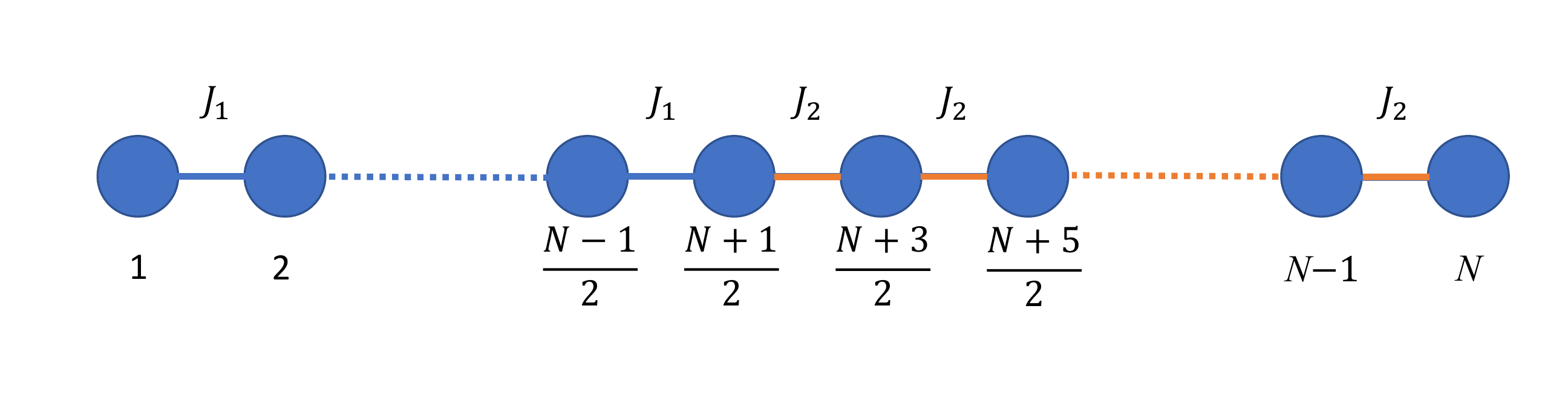}
\caption{Schematic for joint-chain TFIM. Odd system size and open boundary condition are considered. Two spin chain with different nearest neighbour bonds are joined together to form one spin chain. The blue coloured links have exchange interaction strength $J_1$ and the red coloured links have exchange interaction strength $J_2$.}
\label{jointchain_fig}
\end{center}
\end{figure}
\begin{widetext}
We consider the following ansatz to solve equation \eqref{eq:Matrixequation1}
\begin{align}\label{ansatz1}
\phi_{ij}^{T} &= \begin{cases}
A\sin\bigr(k_j(i-1/2)\bigr)+B\cos\bigr(k_j(i-1/2)\bigr) ,&\text{if $1 \leq i \leq \frac{N}{2}$}\\
A_1\sin\bigr(k_j(i-1/2)\bigr)+B_1\cos\bigr(k_j(i-1/2)\bigr) ,&\text{if $(\frac{N}{2}+1) \leq i \leq N $}
\end{cases}
\end{align}
which  implies  the eigenvalues of the form
\begin{eqnarray}
\Lambda_j^2 &= (J_1/h)^2+1-2(J_1/h)\cos(k_j).
\end{eqnarray}
By eliminating all the free parameters in the above ansatz, we obtain 
the expressions for the matrix elements of \{$\phi , \psi$\} in the \{$J_1, J_2, h$\} parameter space as 
\begin{align}
\phi^{T}_{ij} & = \begin{cases}
A\bigl(\frac{J_2}{J_1}\bigr)\frac{\sin\bigr(k_jN/2\bigr)}{\cos\bigl(k_j(N+1/2)\bigr)}\biggl(\frac{\frac{J_1}{h}\sin(k_j(i-1)) - \sin(k_ji)}{\sin\bigl(k_j(N/2+1)\bigr) - \frac{J_1}{h}\sin\bigl(k_jN/2\bigr)}\biggr) , & \text{if $1\leq i \leq\frac{N}{2}$}.\\ \\
A\frac{\sin(k_j(i-N-1)}{\cos\bigl(k_j(N+1/2)\bigr)} ,&\text{if $\frac{N}{2}+1\leq i \leq N$}.
\end{cases} \label{eq:solution1a}\\
\psi_{ij}^{T} &= \begin{cases}
\frac{\frac{J_1}{h}\phi_{i+1j}^{T} - \phi_{ij}^{T}}{\Lambda_j}(1 - \delta_{i,\frac{N}{2}}) + \frac{\frac{J_2}{h}\phi_{\frac{N}{2}+1j}^{T} - \phi_{\frac{N}{2}j}^{T}}{\Lambda_{j}}\delta_{i,\frac{N}{2}} ,&\text{$\Lambda_j \neq 0$,  $ 1 \leq i \leq N $  and  $\phi_{(N+1)j}^{T}\equiv0$} .\\
0 ,&\text{$ \Lambda_j = 0$}.
\end{cases} \label{eq:solution1b}
\end{align} 
Using these expressions in the boundary conditions we get the transcendental equation
\begin{equation}\label{eq:transcendentalequation}
\biggl(\frac{J_2}{J_1}\biggr)^2\sin\biggl(\frac{k_jN}{2}\biggr)\left(\frac{\frac{J_1}{h}\sin\bigl(\frac{k_j(N-2)}{2}\bigr) - \sin\bigl(\frac{k_jN}{2}\bigr)}{\frac{J_1}{h}\sin\bigl(\frac{k_jN}{2}\bigr) - \sin\bigl(\frac{k_j(N+2)}{2}\bigr)}\right) = \sin\biggl(\frac{k_j(N+2)}{2}\biggr) + \biggl(\frac{J_2^2-J_1^2}{hJ_1}\biggr)\sin\biggl(\frac{k_jN}{2}\biggr),
\end{equation}\\
\end{widetext}
where $k_{j}$ is the $j^{th}$ root of \eqref{eq:transcendentalequation}. 
Equation \eqref{eq:transcendentalequation} possesses $N$ roots with no recurrence. 
In some cases, all the roots are real, while in others we witness the emergence of complex roots, but those complex roots are of the form $iu$ or $\pi-iv$, where $u$ and $v$ belong to the real space. All the unique real roots lie in the interval $(0,\pi)$. Next we will explore various correlations (between $\frac{N}{2}$ and $\frac{N}{2}+1$ sites) and transverse magnetisation, which can be evaluated analytically, for a system size $N = 10$. Although in principle with this solution one can study arbitrarily large systems, but for very large system sizes, finding all the $N$ roots of eq. \eqref{eq:transcendentalequation} becomes difficult as it has to be done numerically. 
\section{Model II : Joint chain TFIM} \label{Sec. 4}
Another variant of inhomogeneous TFIM can be considered as model \RomanNumeralCaps 2 (joint-chain transverse field Ising model) which is defined as
\begin{equation}
J_i = \begin{cases}\frac{J_1}{h} ,&\text{for $1 \leq i \leq \frac{N-1}{2}$},\\
\frac{J_2}{h} ,&\text{for $ \frac{N+1}{2} \leq i \leq N-1$}.
\end{cases}
\end{equation}
In this model half of the spin-chain couples with a coupling parameter $J_1$ and the rest half couples with a coupling parameter $J_2$ (See fig. \ref{jointchain_fig} ). After making appropriate changes in the $\alpha$ and $\beta$ matrices the equations correspond to the Hamiltonian of this particular system. The component form of the matrix eq.\eqref{eq:Matrixequation1} is as follows 
\begin{equation}
\phi_{(i-1)j}^{T} + \phi_{(i+1)j}^{T} = \biggr(\frac{J_1^2+h^2-h^2\Lambda_j^2}{hJ_1}\biggr)\phi_{ij}^{T}, 
\end{equation}
for $2 \leq i \leq \left(\frac{N-3}{2}\right)$ and\\ 
\begin{equation}
\phi_{(i-1)j}^{T} + \phi_{(i+1)j}^{T} = \biggr(\frac{J_2^2+h^2-h^2\Lambda_j^2}{hJ_2}\biggr)\phi_{ij}^{T},
\end{equation}\\
for $\left(\frac{N+3}{2}\right) \leq i \leq (N-1)$. \\ \\
The initial condition for this problem is
\begin{align}
\phi_{2j}^{T} &= \biggr(\frac{h^2-h^2\Lambda_j^2}{hJ_1}\biggr)\phi_{1j}^{T}
\end{align}

and the boundary conditions are 
\begin{align}
\phi_{(\frac{N-3}{2})j}^{T} + \phi_{(\frac{N+1}{2})j}^{T} &= \biggr(\frac{J_1^2+h^2-h^2\Lambda_j^2}{hJ_1}\biggr)\phi_{(\frac{N-1}{2})j}^{T},\\
\phi_{(\frac{N-1}{2})j} ^{T}+ \frac{J_2}{J_1}\phi_{(\frac{N+3}{2})j}^{T} &= \biggr(\frac{J_1^2+h^2-h^2\Lambda_j^2}{hJ_1}\biggr)\phi_{(\frac{N+1}{2})j}^{T},\\
\phi_{(N-1)j}^{T} &= \biggr(\frac{J_2^2+h^2-h^2\Lambda_j^2}{hJ_2}\biggr)\phi_{Nj}^{T}.
\end{align} 
\begin{widetext}
We consider the following ansatz to solve this problem
\begin{align}\label{ansatz2}
\phi_{ij}^{T} &= \begin{cases}
A\cos\left(k_j^1(i-1/2)\right) + B\sin\left(k_j^1(i-1/2)\right) ,&\text{if $ 1 \leq i \leq (\frac{N-1}{2})$}\\
A_1\cos\left(k_j^2(i-1/2)\right) + B_1\sin\left(k_j^2(i-1/2)\right) ,&\text{if $ (\frac{N+1}{2}) \leq i \leq N$}
\end{cases} 
\end{align}
which  implies  the eigenvalues of the form
\begin{eqnarray}
\Lambda_j^2 &=(J_1/h)^2+1-2(J_1/h)\cos(k_j^1) = (J_2/h)^2+1-2(J_2/h)\cos(k_j^2).
\end{eqnarray}
It is slightly modified in order to make sure that it works out for this case. The modification is that, instead of having a single quasi-momentum mode $k_j$, here we have two quasi-momentum modes $(k_j^1,k_j^2)$ involved in the ansatz. By eliminating all the free parameters, we get the equations for the matrix elements of \{$\phi , \psi$\} in \{$J_1 , J_2, h$\} parameter space as
\begin{align}
\phi_{ij}^{T} &= \begin{cases}
A\frac{\sin\bigr(k_j^2(N+1)/2\bigr)\bigr(\sin(k^1_ji)-\frac{J_1}{h}\sin(k^1_j(i-1))\bigr)}{\sin\bigr(k_j^2(N+1/2)\bigr)\bigr(\sin\bigr((k_j^1(N+1)/2\bigr) - \frac{J_1}{h}\sin\bigr(k_j^1(N-1)/2\bigr)\bigr)} ,&\text{if $ 1 \leq i \leq (\frac{N-1}{2})$}\\ \\
A\frac{\sin(k^2_j(N+1-i))}{\sin\bigr(k_j^2(N+1/2)\bigr)} ,&\text{ if $\frac{N+1}{2} \leq i \leq N$}
\end{cases}\label{eq:solution2a} \\
\psi_{ij}^{T} &= \begin{cases}
\frac{\frac{J_1}{h}\phi_{(i+1)j}^{T} - \phi_{ij}^{T}}{\Lambda_j} ,&\text{ $1 \leq i \leq \frac{N-2}{2}$ and $\Lambda_j \neq 0 $} \\
\frac{\frac{J_1}{h}\phi_{\frac{N+1}{2}j}^{T} - \phi_{\frac{N-1}{2}j}^{T}}{\Lambda_j} ,&\text{ $ i = \frac{N-1}{2}$ and $\Lambda_j \neq 0 $}\\
\frac{\frac{J_2}{h}\phi_{(i+1)j}^{T} - \phi_{ij}^{T}}{\Lambda_j} ,&\text{ $\frac{N+1}{2} \leq i \leq N, \phi_{(N+1)j}^{T} \equiv 0$ and $\Lambda_j \neq 0 $}\\
0 ,&\text{ if $\Lambda_j = 0$}
\end{cases} \label{eq:solution2b}
\end{align}
and two transcendental equations
\begin{eqnarray}
\frac{J_2^2-J_1^2}{2} &= (J_2h\cos(k^2_j)-J_1h\cos(k^1_j)),\label{eq:trans1}
\end{eqnarray}
\begin{equation}\label{eq:trans2}
\frac{J_2}{J_1}\sin\biggr(\frac{k^2_j(N-1)}{2}\biggr)\biggr(\sin\biggr(\frac{k^1_j(N+1)}{2}\biggr)-\frac{J_1}{h}\sin\biggr(\frac{k^1_j(N-1)}{2}\biggr)\biggr) = \sin\biggr(\frac{k^2_j(N+1)}{2}\biggr)\biggr(\sin\biggr(\frac{k^1_j(N+3)}{2}\biggr)-\frac{J_1}{h}\sin\biggr(\frac{k^1_j(N+1)}{2}\biggr)\biggr),
\end{equation}
\end{widetext}
where $k^1_j$ is the $j^{th}$ root of the coupled transcendental equations \eqref{eq:trans1} and \eqref{eq:trans2} which is paired up with $k^2_j$.
Unlike the previous model, here we have to solve a set of coupled transcendental equations. \eqref{eq:trans1} and \eqref{eq:trans2} in order to find all the allowed quasi-momentum modes. There are total $2N$ solutions to this coupled set of the transcendental equations. $N$ solutions for $k^{1}_j$ modes and $N$ solutions for $k^{2}_j$ modes. Even these equations show the emergence of complex roots but they are either of the form $iu$ or $\pi-iv$, where $u$ and $v$ belong to the real space. All the unique real roots lie in the interval $(0,\pi)$. In this paper we present the correlation evaluated between the junction sites $\frac{N-1}{2}$ and $\frac{N+1}{2}$ and transverse magnetisation for a system size of $N=9$. Here as well, the transcendental equations have to be solved numerically. Hence it becomes difficult to find all $2N$ roots of the transcendental equations \eqref{eq:trans1} and \eqref{eq:trans2}.
\section{Solution using Transfer Matrix Technique} \label{Sec. 6}
The ansatz we have considered to solve the equations presented in Sec. \ref{Sec. 3} and Sec. \ref{Sec. 4}, can be systematically obtained using the transfer matrix \cite{Strand,PhysRevB.41.2173,Peschel2012}that governs the equation 
\begin{equation}
\phi_{(i-1)k}^T+\phi_{(i+1)k}^T = E_k\phi_{ik}^T. \label{eqn:reccur}
\end{equation}
In matrix form this equation can be presented as follows
\begin{equation}
T\Phi_{ik} = \Phi_{(i+1)k}, \label{eqn:reccur1} 
\end{equation} \\
for $2 \leq i \leq (N-1)$.\\ \\
$T$ is the transfer matrix defined by the equation
\begin{equation}
T = \left[\begin{array}{cc}E_k & -1 \\1 & 0\end{array}\right], 
\end{equation} \\
and $\Phi_{ik}$ is defined as 
\begin{equation}
\Phi_{ik} = \left[\begin{array}{c}\phi_{ik}^T \\\\\phi_{(i-1)k}^T\end{array}\right].
\end{equation}
So the general solution to the eq.\eqref{eqn:reccur1} is as follows
\begin{equation}
T^{(p-2)}\Phi_2 = \Phi_{pk} \label{eqn:reccur2}
\end{equation}\\
for  $2 \leq p \leq N$.\\  \\
Let 
\begin{equation}\label{eqn:reccur3}
\Phi_2 = A\left[\begin{array}{c}\exp(\iota3k/2) \\\exp(k\iota/2)\end{array}\right] + B\left[\begin{array}{c}\exp(-\iota3k/2) \\\exp(-k\iota/2)\end{array}\right]
\end{equation}\\ \\
where$\left[\begin{array}{c}\exp(\iota3k/2) \\\exp(k\iota/2)\end{array}\right]$ and
$\left[\begin{array}{c}\exp(-\iota3k/2) \\\exp(-k\iota/2)\end{array}\right]$ are the eigenvectors of $T$ with the eigenvalues $\exp(k\iota)$ and $\exp(-k\iota)$ respectively. \\ \\
$\Rightarrow E_k = 2\cos(k)$. \\ \\
Hence with eqs. \eqref{eqn:reccur2} and  we get \eqref{eqn:reccur3}
\begin{align*}
\Phi_p &= A'\left[\begin{array}{c}\cos(k(2p-1)/2) \\\cos(k(2p-3)/2)\end{array}\right] + B' \left[\begin{array}{c}\sin(k(2p-1)/2) \\\sin(k(2p-3)/2)\end{array}\right]\\ \\
\Rightarrow\phi_{ik}^T &= A'\cos(k(2i-1)/2)+B'\sin(k(2i-1)/2)
\end{align*}
where $A'$ and $B'$ are fixed by the initial and boundary conditions imposed on the eq. \eqref{eqn:reccur}.  As a result we realize that the ansatz we constructed, is connected to a systematic method like the transfer matrix technique.

\section{Signatures of quantum phase transition in inhomogeneous transverse field Ising models}\label{Sec. 7}
\begin{figure*}[htbp]
\begin{center}
\includegraphics[width = \linewidth]{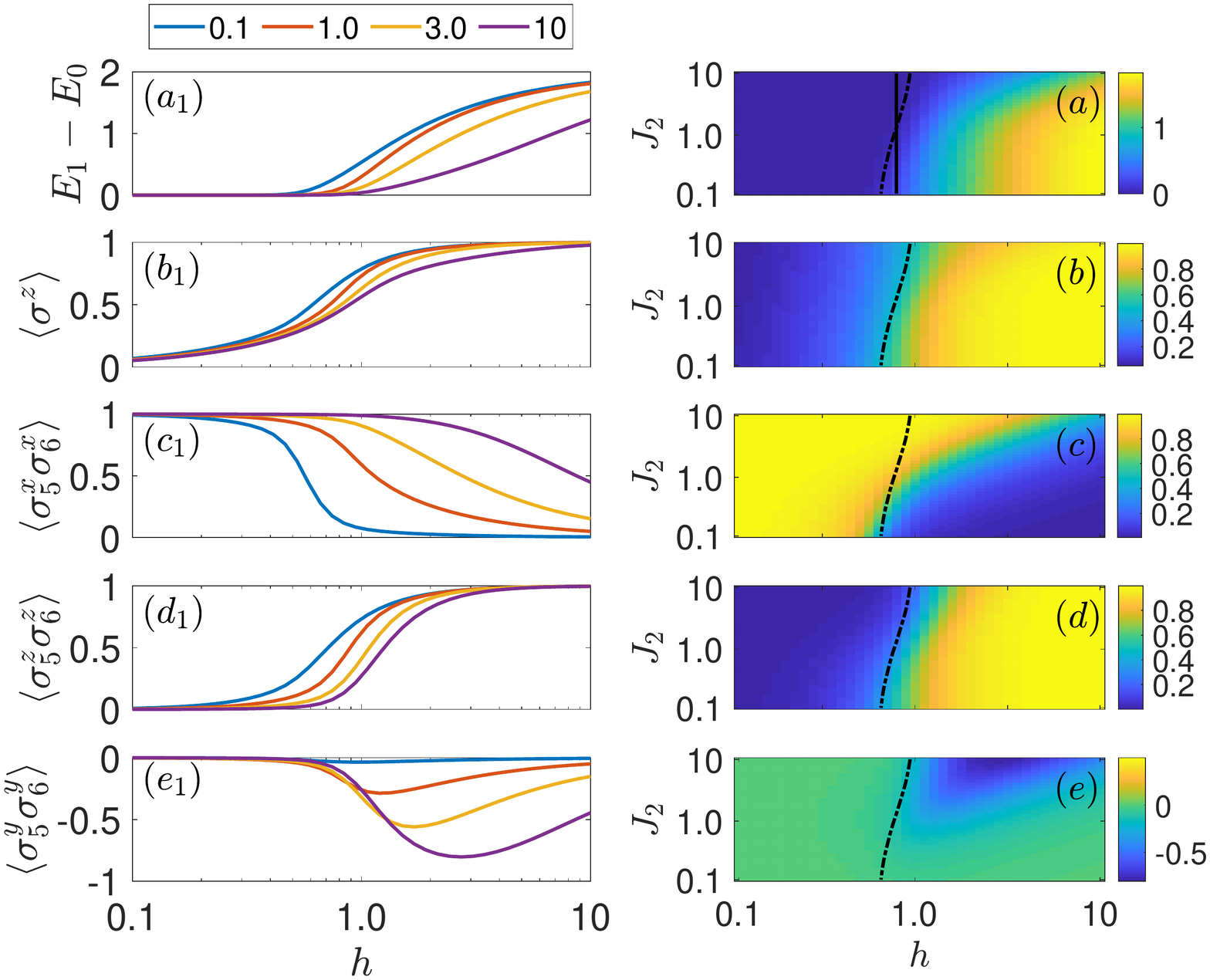}
\caption{All the plots mentioned above are obtained for model \RomanNumeralCaps 1, for system size $N=10$. ($a$) Gap between ground state and first excited state. ($a_1$) Line plots of the gap between the first excited state and ground state for different inhomogeneity coupling strength.  As the inhomogeneity coupling strength increases, the point where the gap between the first excited state and ground state becomes non-zero, keeps shifting ahead along x axis, implying that the transverse magnetic field required for lifting the degeneracy keeps increasing as the inhomogeneity coupling strength increases. ($b$) Transverse magnetization. ($b_1$) Line plots of the transverse magnetization for different inhomogeneity strength coupling. The transverse magnetisation takes longer to reach the value 0.5 as the inhomogeneity interaction strength increases. ($c$) Longitudinal x-x correlation, ($d$) transverse correlation and ($e$) longitudinal y-y correlation between the impurity sites 5 and 6. ($c_1$) , ($d_1$) and ($e_1$) represent the line plots for longitudinal x-x correlation, transverse correlation and longitudinal y-y correlation respectively for different inhomogeneity coupling strengths. The transverse correlation qualitatively shows a very similar behavior as transverse magnetisation. This is because of the fact that the increasing magnetic field tends to align the spins along transverse direction. Hence both transverse correlation and transverse magnetisation increases as the field strength increases. The plots on the right hand side have logarithmic x and y axes where as the for the plots on left hand side, only the x axis is logarithmic, y axis is linear. $J_2$ represents the inhomogeneity coupling strength and $h$ represents the strength of transverse magnetic field at each site. For all the plots, x axis represents transverse magnetic field $h$.}
\label{Fig. 3}
\end{center}
\end{figure*}
\begin{figure*}
\begin{center}
\includegraphics[width =\linewidth]{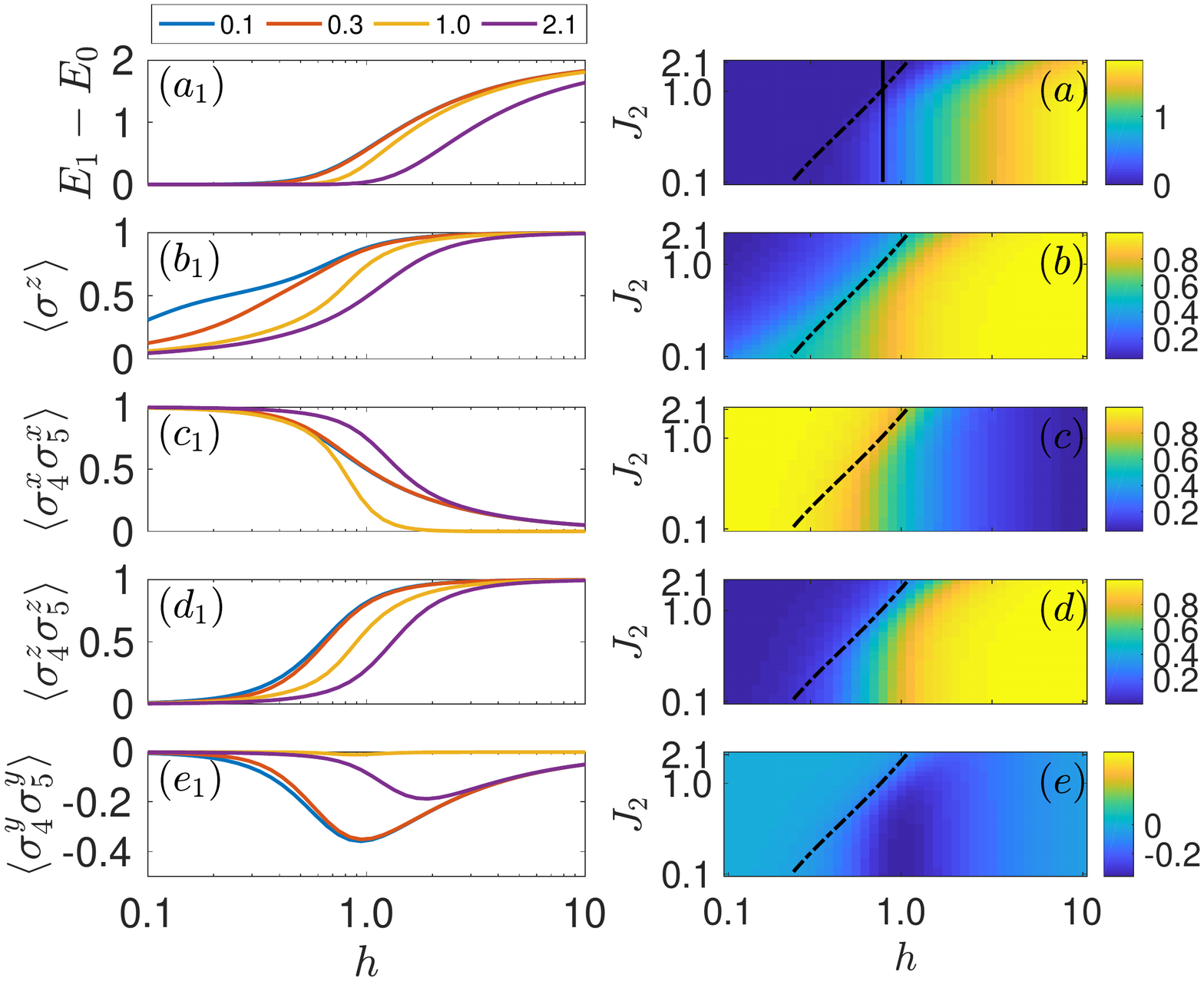}
\caption{These plots are obtained for the model \RomanNumeralCaps 2, for system size $N=9$.
($a$) Gap between ground state and first excited state. ($a_1$) Line plots of the gap between the first excited state and ground state for different inhomogeneity coupling strength. ($b$) Transverse magnetization. ($b_1$) Line plots of the transverse magnetization for different inhomogeneity strength coupling. ($c$) Longitudinal x-x correlation, ($d$) transverse correlation and ($e$) longitudinal y-y correlation between the impurity sites 5 and 6. ($c_1$) , ($d_1$) and ($e_1$) represent the line plots for longitudinal x-x correlation, transverse correlation and longitudinal y-y correlation respectively for different inhomogeneity coupling strengths. Qualitatively all the quantities behave in a similar manner as observed in Fig. \ref{Fig. 3} . The plots on the right hand side have logarithmic x and y axes where as the for the plots on left hand side, only the x axis is logarithmic, y axis is linear. $J_2$ represents the inhomogeneity coupling strength and $h$ represents the strength of transverse magnetic field at each site. For all the plots, x axis represents transverse magnetic field $h$.}
\label{Fig4}
\end{center}
\end{figure*}
The quantum phase transition in TFIM, a canonical system is well understood\cite{Pfeuty1970,Physics1993}. Here the phase transition in TFIM occurs because of the competition between the coupling parameter ($J$) and the transverse magnetic field ($h$). The coupling parameter tends to increase the order in the ground state and the transverse magnetic field tends to disrupt it. So when the coupling strength is too high, the order in the ground state persists and the system is in a ferromagnetic phase, however, when the transverse magnetic field dominates the coupling parameter, the system's ground state becomes becomes a paramagnet. As a result, when the coupling strength is comparable to the transverse magnetic field, we observe a phase transition form an ordered phase to a disordered phase, even at $T = 0K$. 
\\ \\
The phase transition is accompanied by spontaneous symmetry breaking and the system in quantum ferromagnet phase assumes a particular superposition $|0^+\rangle$ or $|0^-\rangle$ ($U_{PF}|0^+\rangle = |0^-\rangle$) as the ground state. In $h\rightarrow0$ limit $|0^+\rangle\approx|\cdots\leftarrow\leftarrow\leftarrow\leftarrow\cdots\rangle$ and $|0^-\rangle\approx|\cdots\rightarrow\rightarrow\rightarrow\rightarrow\cdots\rangle$. As a result the ground state develops spontaneous longitudinal magnetisation $\langle\sigma^x\rangle$.  In $h\rightarrow\infty$ limit, $|0\rangle\approx|\cdots\uparrow\uparrow\uparrow\cdots\rangle$. Hence all the spontaneous longitudinal magnetisation is lost. Thus, longitudinal magnetisation serves as an order parameter across the phase transition. However, for finite sized systems, $\langle\sigma^x\rangle$, when expressed in terms of Jordan-Wigner fermionic operators gives a non-local expression which is difficult to evaluate\cite{Osborne2002,Heyl2018}. Hence, we measure other quantities which have similar behavior as $\langle\sigma^x\rangle$, across the phase transition like transverse magnetisation $\langle\sigma^z\rangle$.
\\ \\
 In inhomogeneous TFIM, we have two free parameters ($J_2$ and $h$). The background coupling strength ($J_1$) is normalised to 1 so that we can study the effects of inhomogeneity coupling strength ($J_2$) with respect to this scale. Analyzing the effects of the inhomogeneity coupling on the phase transition line can tell us a lot about the occurrence of the quantum phase transition in these models. Using the ansatz \eqref{ansatz1} and \eqref{ansatz2}, we solve the matrix equations that characterize model \RomanNumeralCaps 1 and model \RomanNumeralCaps 2 respectively. We use the $\phi$ and $\psi$ matrices, that we obtain as solutions of those matrix equations, to calculate different physical quantities and explore the effect of adding an inhomogeneity in the system. The phase transition in these models is accompanied by symmetry breaking (Global phase flip symmetry), which results in a change in the ground state degeneracy structure across the phase transition. This can be explicitly checked by calculating the gap between the ground state and the first excited state $(E_1-E_0)$ which we show in fig. \ref{Fig. 3} ($a$), ($a_1$) and fig. \ref{Fig4} ($a$), ($a_1$). We can get an approximate estimate of the phase transition line by locating the values of parameters ($J_2$ and $h$) where the transverse magnetisation  $\langle\sigma^z\rangle$ is 0.5. We show this line, in fig. \ref{Fig. 3} ($a$) and fig. \ref{Fig4} ($a$) by the dashed curve. The solid black line represents the phase transition line in fig. \ref{Fig. 3} ($a$) and fig. \ref{Fig4} ($a$), for TFIM. The degenerate ground state of inhomogeneous TFIM in the quantum ferromagnet phase has very different properties than the non-degenerate ground state, in the quantum paramagnet phase. We expect the region of parameter space within which this rapid transition takes place for finite sized systems, keeps getting narrower as the system size tends to infinity. In the end it collapses with the boundary line of the no-gap region. 
 We can calculate two site site correlation function between sites $i$ and $j$ in x-x direction  $\langle\sigma_i^x\sigma_j^x\rangle$, y-y direction
 $\langle\sigma_i^y\sigma_j^y\rangle$ and z-z direction $\langle\sigma_i^z\sigma_j^z\rangle$ as following:
  \begin{align}
  \langle\sigma_i^x\sigma_j^x\rangle &= \left|\begin{array}{cccc}G_{i,i+1} & G_{i,i+2} & \cdots & G_{i,j} \\\vdots &  &  &  \\G_{j-1,i+1} & \cdots & \cdots & G_{j-1,j}\end{array}\right|, \\ 
  \langle\sigma_i^y\sigma_j^y\rangle &= \left|\begin{array}{cccc}G_{i+1,i} & G_{i+1,i+1} & \cdots & G_{i+1,j-1} \\\vdots &  &  & \vdots \\G_{j,i} & \cdots & \cdots & G_{j,j-1}\end{array}\right|, \\
  \langle\sigma_i^z\sigma_j^z\rangle &= \left|\begin{array}{cc}G_{i,i} & G_{j,i} \\G_{i,j} & G_{j,j}\end{array}\right|, 
\end{align}  
where $G_{i,j} = -\left(\psi^T\phi\right)_{i,j}$.\cite{Lieb1961,Apollaro2017}
The transverse magnetization per site is given by 
  \begin{align}
  \langle\sigma_i^z\rangle &= 1-2\sum_{q=1}^{N}\left(\frac{\psi_{qi}-\phi_{qi}}{2},\right)^2
\end{align}
and the total transverse magnetization as
\begin{eqnarray}
    \langle\sigma^z\rangle &= \frac{\sum_{i=1}^{N} \langle\sigma_i^z\rangle}{N}.
\end{eqnarray}
We use the exact solutions for model  \RomanNumeralCaps 1 and \RomanNumeralCaps 2  to compute the two site correlation functions and transverse magnetisation $\langle\sigma^z\rangle$ defined in the above equations. We show some signatures of quantum phase transition in the ground state of these models. In particular the transverse magnetisation $\langle\sigma^z\rangle$ and transverse correlation $\langle\sigma^z_i\sigma^z_j\rangle$ are observed to be highly sensitive to this phase transition. The correlation function $\langle\sigma^z_i\sigma^z_j\rangle$ and $|\langle\sigma^z\rangle|$ go from 0 to 1 as $h$ goes from 0 to infinity, for any value of $J_2$ and for all $(i,j)$ in a given spin chain. We measure the transverse magnetisation $\langle\sigma^z\rangle$ and show it in fig. \ref{Fig. 3} ($b$), ($b_1$) and fig. \ref{Fig4} ($b$), ($b_1$), as a function of $h$ for different values of $J_2$. This quantity varies smoothly from 0 to 1 with respect to $h$, for a finite system. For a given value of $J_2$, this gap is a smooth function of $h$. The longitudinal x-x correlation $\langle\sigma^x_i\sigma^x_j\rangle$  qualitatively behaves in an exactly opposite manner to $\langle\sigma^z_i\sigma^z_j\rangle$, for any value of $J_2$ and for all $(i,j)$. When the transverse magnetic field is low, all the spins tend to point in the same direction along x axis,  however, increasing transverse magnetic field, tends to align the spins along z axis (This represents the domain breakdown). As a consequence, increasing the transverse magnetic field results in reduction of longitudinal x-x correlations and increase in transverse correlations. This behavior is well depicted in the fig. \ref{Fig. 3} ($c$), ($c_1$), ($d$), ($d_1$) and fig. \ref{Fig4} ($c$), ($c_1$), ($d$), ($d_1$). In the limiting case $h\rightarrow0$, for any value of $J_2$, the ground state of the system is $|\cdots\leftarrow\leftarrow\leftarrow\leftarrow\cdots\rangle$ and in the limit $h\rightarrow\infty$ the ground state is $|\uparrow\uparrow\uparrow\cdots\uparrow\rangle$. Here we consider the symmetry broken ground state, out of the entire degenerate space. This does not result in loss of generality, as these correlation functions and  transverse magnetization are insensitive to the choice of superposition of the symmetry broken ground states that we consider. As a result, because of the analytic behavior of correlation functions and transverse magnetization, we can qualitatively say that there has to be a region in the parameter space where these quantities undergo a change, inherently indicating the loss of degeneracy from the ground state of the system, for finite sized systems. This is observed in fig. \ref{Fig. 3} and fig. \ref{Fig4}. On the contrary $\langle\sigma^y_i\sigma^y_j\rangle$ correlation function tends to be 0 in both the cases for all $(i,j)$, which is again a consequence of the limiting case behavior of the ground state. This is well depicted in the fig. \ref{Fig. 3} ($e$), ($e_1$) and fig. \ref{Fig4} ($e$), ($e_1$).
\\ \\
We conclude this section with the following observation we make in the exactly solvable models \RomanNumeralCaps 1 and \RomanNumeralCaps 2 that, when the inhomogeneity coupling strength $J_2 > 1$, we see that the occurrence of quantum phase transition is delayed as compared to TFIM. On the contrary for $J_2 < 1$ we see that the occurrence of the phase transition is preponed. This behavior is expected based on a simple mean field argument. The phase transition can be thought of as arising due to a competition between the Ising terms ($J's$) and the transverse magnetic field ($h$). In a mean field picture the competition is between average of coupling terms ($J's$)  with transverse magnetisation ($h$). As $J_2$ is decreased, the average $J$ decreases as well requiring smaller $h$ for phase transition and vice versa. 

\section{Conclusion} \label{Sec. 8}
We have studied the QPT which occurs in the models, namely impurity based transverse field Ising model and joint-chain Ising model. We spotted the effects of introducing inhomogeneities on the phase transition using the exact solutions that we obtained in the body of this paper. To do so, we use the ansatz that we construct and analytically diagonalise the two systems which are taken into consideration. Also we give a systematic way  of constructing this ansatz using the transfer matrix technique. This solution can easily be generalised to multiple impurity site problems (impurities refers to bond defects), where the impurities should be separated by at least two lattice sites on the one dimensional lattice. The solution for joint-chain Ising model can be extended to multiple junction problems, in which more than two Ising chains with different coupling strengths are connected together. These junctions should be separated by at-least two lattice points. Using the analysis employed in this paper, we can extend the analytical solutions for problems which have junctions as whell as isolated impurities too. Also we observe that the inhomogeneity coupling strength acts as a tuneable parameter which controls the occurrence of QPT in these models. The quench dynamics of these models holds a lot of promise for the future development of models useful in the implementation quantum information protocols.

\bibliographystyle{apsrev4-1}
\bibliography{Paper1}

\end{document}